\documentclass[journal,12pt,draftclsnofoot,onecolumn,twoside]{IEEEtran}
\usepackage{cite}

\ifCLASSINFOpdf
\usepackage[pdftex]{graphicx}
\else
\usepackage[dvips]{graphicx}
\fi
\usepackage{epstopdf}					
\graphicspath{{figures/}}

\usepackage[cmex10]{amsmath}
\usepackage{amsfonts}

\usepackage[footnotesize,belowskip=-14pt,aboveskip=0pt]{caption}
\begin{document}
%
\title{On Partially Overlapping Coexistence for Dynamic Spectrum Access in Cognitive Radio}

\author{{Ebrahim Bedeer, 
Mohamed Marey, 
Octavia Dobre, and 
Kareem Baddour \IEEEauthorrefmark{2}}\\
\IEEEauthorblockA{Faculty of Engineering and Applied Science, Memorial University of Newfoundland,
St. John's, NL, Canada\\ 
\IEEEauthorrefmark{2} Communications Research Centre, Ottawa, ON, Canada\\
Email: \{e.bedeer, mmarey, odobre\}@mun.ca, kareem.baddour@crc.ca}
}



\maketitle

\begin{abstract}
In this paper, we study partially overlapping coexistence scenarios in cognitive radio environment. We consider an Orthogonal Frequency Division Multiplexing (OFDM) cognitive system coexisting with a narrow-band (NB) and an OFDM primary system, respectively. We focus on finding the minimum frequency separation between the coexisting systems to meet a certain target BER. Windowing and nulling are used as simple techniques to reduce the OFDM out-of-band radiations, and, hence decrease the separation. The effect of these techniques on the OFDM spectral efficiency and PAPR is also studied. 
\end{abstract}


%
\IEEEpeerreviewmaketitle

\section{Introduction}

With the advent of novel high data rate wireless applications, development of existing wireless services, and emergence of new services, the demand for additional bandwidth is rapidly increasing. Frequency spectrum is allocated in each country by government agencies, which impose regulations on its usage. The vast majority of the available spectrum has been already licensed, and it is becoming increasingly difficult to find spectrum that can be used either to expand the existing services or to introduce new ones. The current regulations do not allow unlicensed access to licensed spectrum, and the unlicensed frequency bands are heavily populated and prone to interference; hence, there is a scarcity of the frequency spectrum. However, recent measurements show that the spectrum utilization is sparse both spatially and temporally \cite{fcc2005notice}. As such, there is practically an underutilization of the frequency spectrum.  ``Spectrum holes'' occur dynamically, depending on the geographical area and time, and can be exploited by unlicensed or cognitive users (CU). In order to utilize these ``spectrum holes'' under dynamically changing environment conditions, which is also referred to as dynamic spectrum access (DSA), a new wireless communication technology is required. Cognitive Radio (CR) has emerged as a solution to DSA, due to its adaptability and reconfigurability \cite{haykin2005cognitive}. Most of the work on DSA has focused on improving spectrum utilization without interfering with primary users (PU) \cite{haykin2005cognitive, tandra2009spectrum}. 

\begin{figure}[!t]
	\centering
  \includegraphics[width=0.50\textwidth]{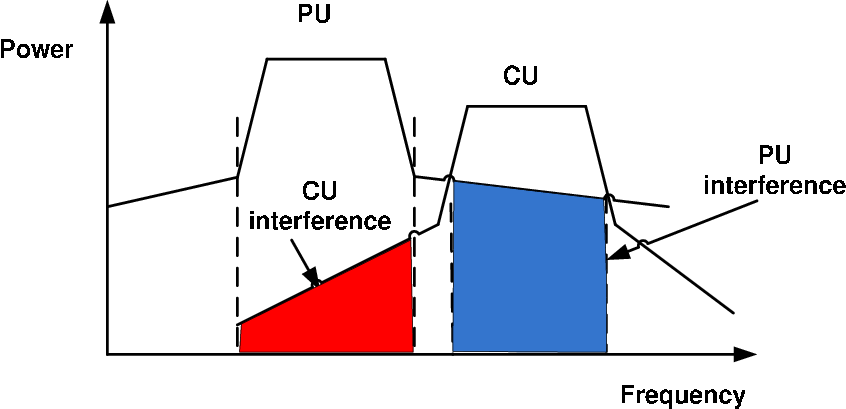}
	\caption{Partially overlapping coexistence concept.}
	\label{fig:Coex}
\end{figure}

An alternative to this approach, which can lead to a more efficient utilization of the frequency spectrum, is to adapt the transmitted waveform to coexist in the same or partially overlapping frequency bands without cooperation. To ensure reliable transmission, such a strategy needs to consider the interference effects on PUs, as well as on CUs. Most studies have focused on the coexistence in the same frequency band (wideband underlying cognitive radio systems) \cite{4802197}. Hence, there is a need to investigate the interference conditions which allow a reliable transmission in partially overlapping coexistence  scenarios. The concept of the partially overlapping is illustrated in Fig. \ref{fig:Coex}. Apparently, it is important to explore techniques to reduce mutual interference of the coexisting systems to preserve the reliability of the transmission, yet achieving a high spectral efficiency. 

OFDM is adopted by many wireless standards, e.g., IEEE 802.11 WLAN, IEEE 802.16 WiMAX, and LTE \cite{4907554}, and considered as an attractive modulation technique for CR \cite{4907554}. Hence, we explore the coexistence between an OFDM CU and a PU system; the latter is either a narrow-band (NB) or another OFDM-based system. The performance of each system is studied in terms of the bit error rate (BER) under diverse coexisting conditions, such as different signal-to-interference ratios (SIRs) and frequency separation between coexisting systems. We focus on finding the minimum frequency separation between the PU and CU to meet a certain average target BER. In addition, we study two methods, namely, windowing and nulling subcarriers to  reduce the CU out-of-band radiation \cite{weiss2005mutual}, thus, decreasing the interference to the PU and increasing the frequency separation.  The effect of these techniques on the CU spectral efficiency and peak-to-average power ratio (PAPR) is also investigated. 

The reminder of this paper is organized as follows: Section \ref{sec:model} describes the system models and the coexistence scenarios. Simulation results are presented in Section \ref{sec:sim}. Finally, conclusions are drawn in Section \ref{sec:conc}.

\section{System Models and Coexistence Scenarios} \label{sec:model}

This section introduces the models of the OFDM and NB uncoordinated systems, as well as the considered coexistence scenarios. Moreover, windowing and nulling subcarriers are presented as simple techniques to reduce the OFDM CU out-of-band radiation. 

\subsection{OFDM system model}

The OFDM signal is generated by mapping the modulated input signal to $N_u$ orthogonal subcarriers. A cyclic prefix (CP) of duration $T_{cp}$ is then inserted in order to mitigate the inter-symbol interference (ISI) caused by the channel delay spread. A postfix of duration $T_p$ is appended to the end of the useful OFDM symbol for  the windowing purpose.  The CP and the postfix are formed by repeating the last $T_{cp}$ part and the first $T_p$ part of the useful OFDM symbol, respectively. Finally, the resultant signal is multiplied by a window to reduce the out-of-band radiation. The resulting baseband signal is \cite{van2000ofdm}
\begin{eqnarray}
s_{\textsc{OFDM}}(t) = \frac{1}{\sqrt{T_o}}\sum_{n = -\infty}^{\infty}\sum_{k \in \Omega }^{} a^n_k e^{i 2\pi f_k (t - n T_o)} w(t - n T_o), 
\label{eq:ofdm_Tx}
\end{eqnarray}
where $\Omega$ represents the set of active/useful subcarriers with cardinality $N_u$, $a^n_k$ is the data symbol transmitted on the $kth$ subcarrier of the $nth$ OFDM symbol, $T_o = T_u + T_{cp} + T_p$ is the OFDM symbol duration, with $T_u$ as the useful OFDM symbol duration, and $w(t)$ is the window function.


Assuming that the OFDM signal propagates over multiple paths, the received OFDM signal is given by \cite{proakis2008digital}
\begin{equation}
r_{\textsc{OFDM}}(t)=\sum_{j=0}^{J-1}h_j(t) s_{\textsc{OFDM}}(t-\tau_j(t)) + n(t),
\label{eq:re_OFDM}
\end{equation}
where $J$ is the number of channel paths, $h_j(t)$ and $\tau_j(t)$ are the gain and the propagation delay of the $jth$ path, respectively, and $n(t)$ is the additive white Gaussian noise (AWGN). At the OFDM receiver, the CP and the postfix are removed, and the resultant signal is demodulated by a bank of demodulators, after which decision on the data symbols is made.

The OFDM CU shapes its spectrum to reduce interference to the PU. Different windows, including Hamming, Hann, Kaiser, Bartlett, and widows satisfying the Nyquist criteria, can be used to reduce the out-of-band radiation \cite{weiss2005mutual}. The commonly used raised cosine window is considered in this paper which is defined as \cite{weiss2005mutual,van2000ofdm}
\begin{eqnarray}
w(t) = \left\{
\begin{array}{ll}
\frac{1}{2} + \frac{1}{2}cos(\pi+\frac{\pi t}{\beta T_s}) & 0 \leq  t < \beta T_s\\
1 & \beta T_s \leq  t < T_s\\
\frac{1}{2} + \frac{1}{2}cos(\pi+\frac{\pi (t-T_s)}{\beta T_s}) & T_s \leq t < (1+\beta)T_s,
\end{array} \right.
\label{eq:window}
\end{eqnarray}
where $\beta$ denotes the roll-off factor, and $T_s =\frac{T_o}{(1+\beta)}$. The other simple technique to reduce the OFDM out-of-band radiation is to deactivate/null subcarriers.

Our interest extends to the effect of these techniques on the OFDM signal spectral efficiency and PAPR. The spectral efficiency, $\zeta$, is defined as the information rate that can be transmitted for a given bandwidth \cite{proakis2008digital}. For an OFDM signal, one can express $\zeta$ as 
\begin{eqnarray}
\zeta = \frac{m N_u /(T_s(1+\beta))}{N \Delta F},
\end{eqnarray}
where $m$ is the number of bits per data symbol, $N$ is the total number of subcarriers, and $\Delta F$ is the subcarrier spacing. It is clearly that $\zeta$ is an increasing function of $N_u$, and a decreasing function of $\beta$. So, both nulling and windowing reduces $\zeta$ by decreasing $N_u$ and increasing $\beta$ respectively. On the other hand, the signal PAPR is \cite{van2000ofdm}
\begin{eqnarray}
PAPR = \frac{max[s(t) s^*(t)]}{\mathbb{E}[s(t) s^*(t)]}, \qquad t \in [0,T_s).
\label{eq:papr}
\end{eqnarray}
Numerical examples of the effect of windowing and subcarrier nulling on the PAPR are provided in Section \ref{sec:sim}.

\subsection{NB system model}
The NB PU received signal, $r_{\textsc{NB}}(t)$, can be written as \cite{marey2007analysis}
\begin{equation}
r_{\textsc{NB}}(t) = \sum_{p =0}^{P-1}h_{p}(t)s_{\textsc{NB}}(t-\tau _{p}(t)) e^{i2\pi f_{c}(t-\tau _{p}(t))},
\label{eq:int}
\end{equation}
where $P$ is the number of channel paths, $h_{p}(t)$ and $\tau _{p}(t)$ are the $pth$ channel path gain and delay, respectively, and $f_{c}$ is the NB PU frequency deviation from the OFDM carrier frequency. $s_{\textsc{NB}}(t)$ is the transmitted NB PU signal, given by $\sum_{k =-\infty }^{\infty}b_{k}p(t-kT-\xi),
$
where $b_{k}$ is the $kth$ data symbol, $p(t)$ is the impulse response of the transmit root raised cosine filter, $\xi$ is the time delay, and $T$ is the symbol period. 

\subsection{Coexistence between NB and OFDM systems}
The signal-to-interference ratio (SIR) defined at the output of the NB PU receiver matched filter is given by $SIR  =  \frac{\rho}{\rho_{I}},
$
where $\rho$ and $\rho_{I}$ are the NB PU signal and the OFDM CU signal average powers at the output of the NB matched filter, respectively. The average power of the NB PU signal at the output of the root-raised cosine matched filter is given by 
\setlength{\arraycolsep}{0.0em}
\begin{eqnarray}
\rho&{}={}&\frac{1}{T}\mathbb{E}\left[\int_{-\infty }^{\infty}\right. \nonumber\\
&&\:\left. \left | \sum_{p = 0}^{P-1} h_{p}(t)\sum_{k = -\infty}^{\infty}b_{k}p(t-kT-\xi-\tau_{p}(t)) \star p(t)\right|^{2}dt\right],\nonumber\\
\end{eqnarray}
\setlength{\arraycolsep}{5pt}
where $\star$ denotes the convolution. By using that $\sum_{p=0}^{P-1} \sum_{p'=0}^{P-1} \mathbb{E}[h_p(t)h_{p'}^*(t)] \neq 0 $ and 

$\sum_{k = - \infty }^{\infty} \sum_{k' = - \infty }^{\infty} \mathbb{E}[b_k b_{k'}^*] \neq 0 $ if and only if $p = p'$ and $k = k'$, respectively, $\sum_{p=0}^{P-1} \mathbb{E}|h_p(t)|^2 = 1$, and that the energy of the root-raised cosine pulse shape equals 1 \cite{proakis2008digital}, and assuming time-independent path delays for the channel, one can easily show that 
\begin{eqnarray}
\rho & = & \sigma_b^2 (1-\frac{\alpha}{4}), \label{eq:PinterFinal}
\end{eqnarray}
where $\sigma_b^2 = \mathbb{E} [\left | b_{k} \right |^2]$ is the power per NB PU data symbol and $\alpha$ is the NB PU root-raised cosine roll-off factor.

The OFDM CU symbol after the NB PU receiver matched filter is expressed as
\setlength{\arraycolsep}{0.0em}
\begin{eqnarray}
s_I(t)&{}={}&\left [\sum_{j=0}^{J-1}h_j(t) s_{\textsc{OFDM}}(t-\tau_j(t)) e^{-i 2\pi f_c(t-\tau_j(t))} \right ] \star p(t),\nonumber\\
\end{eqnarray}
\setlength{\arraycolsep}{5pt} 
and the average power $\rho_{I}$ per OFDM symbol after the NB receiver matched filter can be written as
\setlength{\arraycolsep}{0.0em}
\begin{eqnarray}
\rho_{I}&{}={}&\frac{1}{T_o} \int_{0}^{T} \mathbb{E}[ s_I(t) s_I^*(t)]  \:dt,\nonumber\\
\end{eqnarray}
\setlength{\arraycolsep}{5pt}
where * denotes the complex conjugate. After straightforward mathematical manipulations, the SIR can be shown to be \begin{eqnarray}
SIR & = &  \frac{\sigma_b^2(1-\alpha/4)}{\sigma_a^{2} /T_o} C,
\label{eq:SIR_NB_OFDM}
\end{eqnarray}
where $\sigma_a^2 = \mathbb{E} [\left | a_k \right |^2]$ is the power per OFDM CU data symbol and  
\setlength{\arraycolsep}{0.0em}
\begin{eqnarray}
C&{}={}&\left[\int_{0}^{T} \sum_{k \in \Omega} (e^{-i 2\pi (f_k + f_c) (t-\tau_j(t))} w(t-\tau_j(t))\star p(t))\right.\nonumber\\
&&\:\left.(e^{i 2\pi (f_{k} + f_c) (t-\tau_{j}(t))} w(t-\tau_{j}(t))\star p^*(t)) dt\right]^{-1} \nonumber
\end{eqnarray}
\setlength{\arraycolsep}{5pt}
is a constant evaluated numerically through computer simulations.

\subsection{Coexistence between two OFDM systems}
The SIR defined at the OFDM PU receiver is given by $SIR  =  \frac{\rho}{\rho_I}$,
where, in this case, $\rho$ and $\rho_I$ represent the PU OFDM signal and the CU OFDM signal average powers, respectively. The average power of an OFDM symbol is 
\begin{eqnarray}
\rho  =  \frac{1}{T_o}\mathbb{E}\left[\int_{-\infty }^{\infty}\left | \sum_{j=0}^{J-1}h_j(t) s_{\textsc{OFDM}}(t-\tau_j(t)) \right |^{2}dt\right],
\end{eqnarray}
and, after direct mathematical manipulation, one can show that $\rho$ can be expressed as
\begin{eqnarray}
\rho & = & \frac{1-\frac{1}{4}\beta}{1+\beta} N_u \sigma_a^2.
\label{eq:hh}
\end{eqnarray}
Hence, the SIR is given by 
\begin{eqnarray}
SIR = \frac{\frac{1 - \frac{1}{4}\beta}{1+\beta}N_u \sigma_a^2}{\frac{1 - \frac{1}{4}\beta'}{1+\beta'}N_u^{'}\sigma_a^{'2}},
\end{eqnarray}
where $(.)'$ represents the PU OFDM system parameters.

\section{Simulation and Numerical Results} \label{sec:sim}
\subsection{Simulation Setup}

The parameters of the systems considered in this study are provided in Table \ref{tab:SimPar}. The time delay $\xi$ is a random variable uniformly distributed between 0 and the PU symbol duration. AWGN and fading channels are considered. A frequency selective fading channel is used with the OFDM, while a frequency non-selective fading channel is used with the NB. An exponential power delay profile is considered for the frequency selective channel \cite{morelli2004timing}, with the average received power for the $jth$ path equal to $E_h e^{-j\Xi}$, $j$ = 0, ..., 4, where $E_h$ is a constant chosen such that the average energy per subcarrier is normalized to unity, and $\Xi$ represents the decay factor, $\Xi = \frac{1}{5}$. On the other hand, for the frequency non-selective channel, the transmitted signal is multiplied by a complex Gaussian random variable with zero-mean and variance equal to one. The target BER is selected to be $10^{-4}$. The normalized location, $F_n$, of the PU is defined as the difference between the PU center frequency and the OFDM CU edge frequency normalized to the subcarrier spacing $\Delta F$, $F_n$ = $(f_c-BW)/\Delta F$ ($BW$ is the OFDM CU bandwidth), i.e., $F_n$ = 10 means that the PU is located 10 subcarriers away from the OFDM CU edge frequency.
\begin{table}[!t]
  \centering
  \caption{OFDM and NB system parameters.}
    \begin{tabular}{rl}
    \hline
    {\bf } & {\bf OFDM system} \\ \hline
    { Bandwidth, $BW$} & 1.25 MHz \\
    { Number of subcarriers, $N$} & 128 \\
    { Subcarrier spacing, $\Delta F$} & 9.7656 kHz \\
    { Useful symbol duration, $T_u$} & 102.4 $\mu$sec \\
    { CP duration, $T_{cp}$} & 0.25$T_u$ = 25.6 $\mu$sec \\
    { Modulation type} & Quadrature PSK (B/QPSK) \\ \hline
    {\bf } & {\bf NB system} \\ \hline
    { Bandwidth, $BW_N$} & 15 kHz \\
    { Roll-off factor, $\alpha$} & 0.35 \\
    { Modulation type} & QPSK \\ 
    \hline
    \end{tabular}
  \label{tab:SimPar}
\end{table}

\subsection{Coexistence between OFDM and NB systems}

\begin{figure}[!t]
	\centering
		\includegraphics[width=0.50\textwidth]{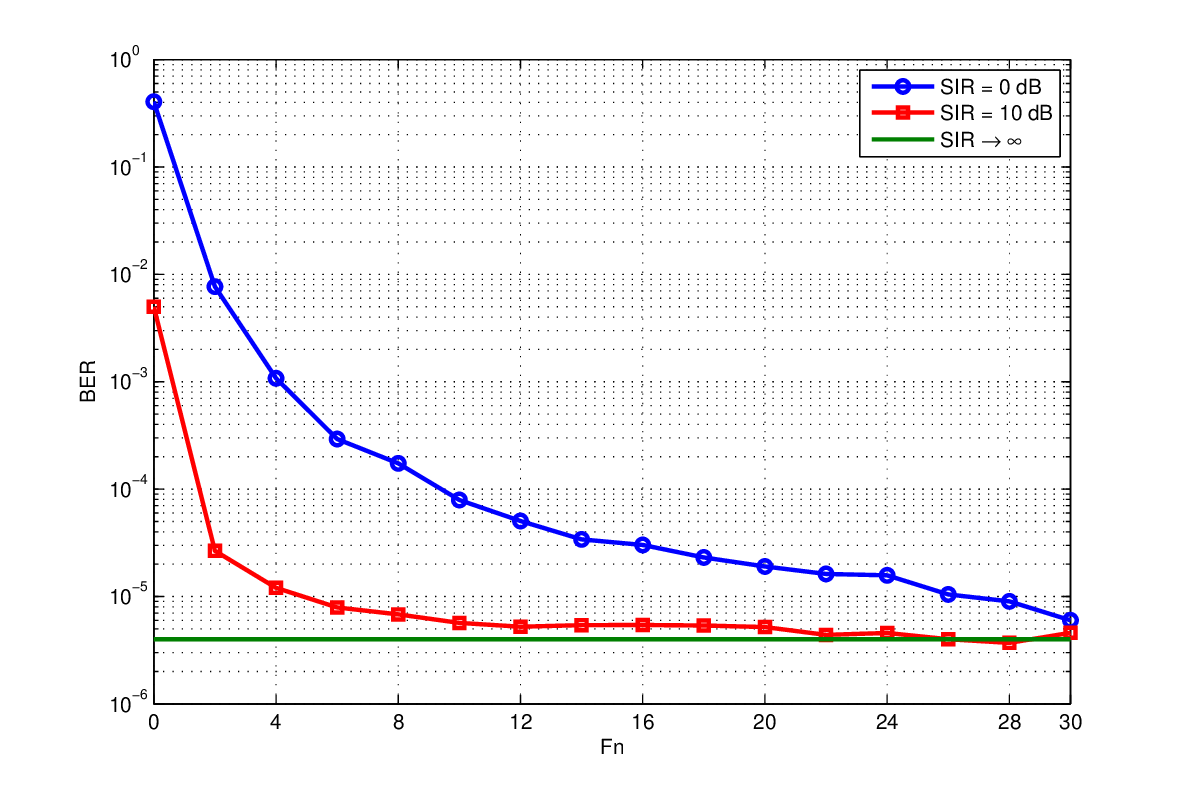}
	\caption{NB PU BER as a function of $F_n$ in AWGN channel at $\frac{E_b}{N_o} = 10 \: dB$.}
	\label{fig:Results_NBI_OFDM_vs_F}
\end{figure}

Fig. \ref{fig:Results_NBI_OFDM_vs_F} illustrates the effect of the OFDM CU location and SIR on the NB PU BER in AWGN channel. As one can notice, increasing the value of $F_n$ or SIR improves the BER performance of the PU till it reaches an asymptotic value that depends only on $\frac{E_b}{N_o}$; this asymptotic value is given by $Q\left (  \sqrt{\frac{2E_b}{N_o}} \right )$ with $Q(.)$ as the Q-function \cite{proakis2008digital}. Moreover, the minimum frequency separation required to reach the target BER depends on SIR. For example, at SIR = 0 dB, this  separation nearly equals $10\Delta F$, while at SIR = 10 dB only $1.5\Delta F$ is needed to obtain the same performance.

\begin{figure}[!t]
	\centering
		\includegraphics[width=0.50\textwidth]{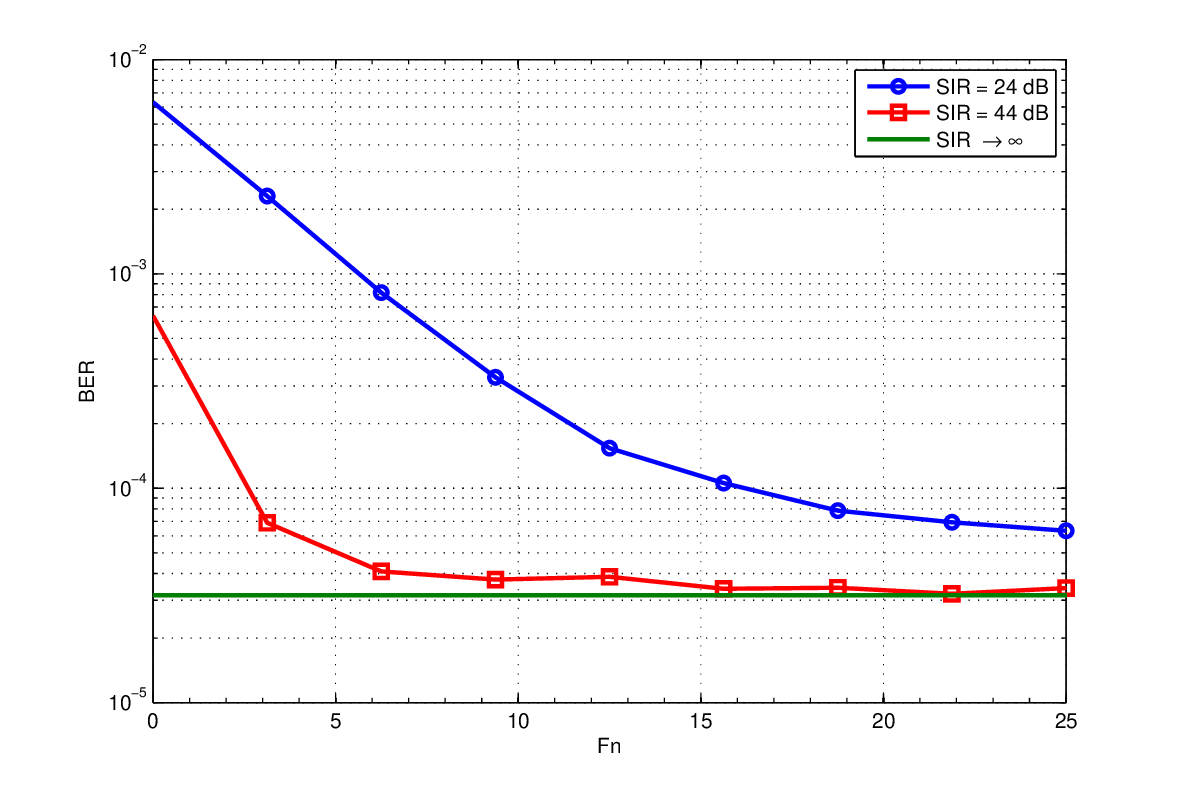}
	\caption{OFDM CU BER as a function of $F_n$ in AWGN channel at $\frac{E_b}{N_o} = 10 \: dB$.}
	\label{fig:Results_NBI_vs_F}
\end{figure}

On the other hand, the effect of the NB PU location and power on the OFDM CU BER is shown in Fig. \ref{fig:Results_NBI_vs_F}. The SIR values that the OFDM CU system experiences, which correspond to the 0 dB and 10 dB SIR at the NB PU, are 24 dB and 44 dB, respectively. As in the previous case, the minimum frequency separation depends on the SIR value. To meet the target BER for the OFDM CU system, this separation equals $16\Delta F$ at SIR = 24 dB and $2.5\Delta F$ at SIR = 44 dB. Hence, for both systems to coexist and meet the target BER, the minimum separation should be $16\Delta F$ for SIR = 0 dB (24 dB) and $2.5\Delta F$ for SIR = 10 dB (44 dB), being actually imposed by the target performance of the OFDM CU system.

\begin{figure}[!t]
	\centering
		\includegraphics[width=0.50\textwidth]{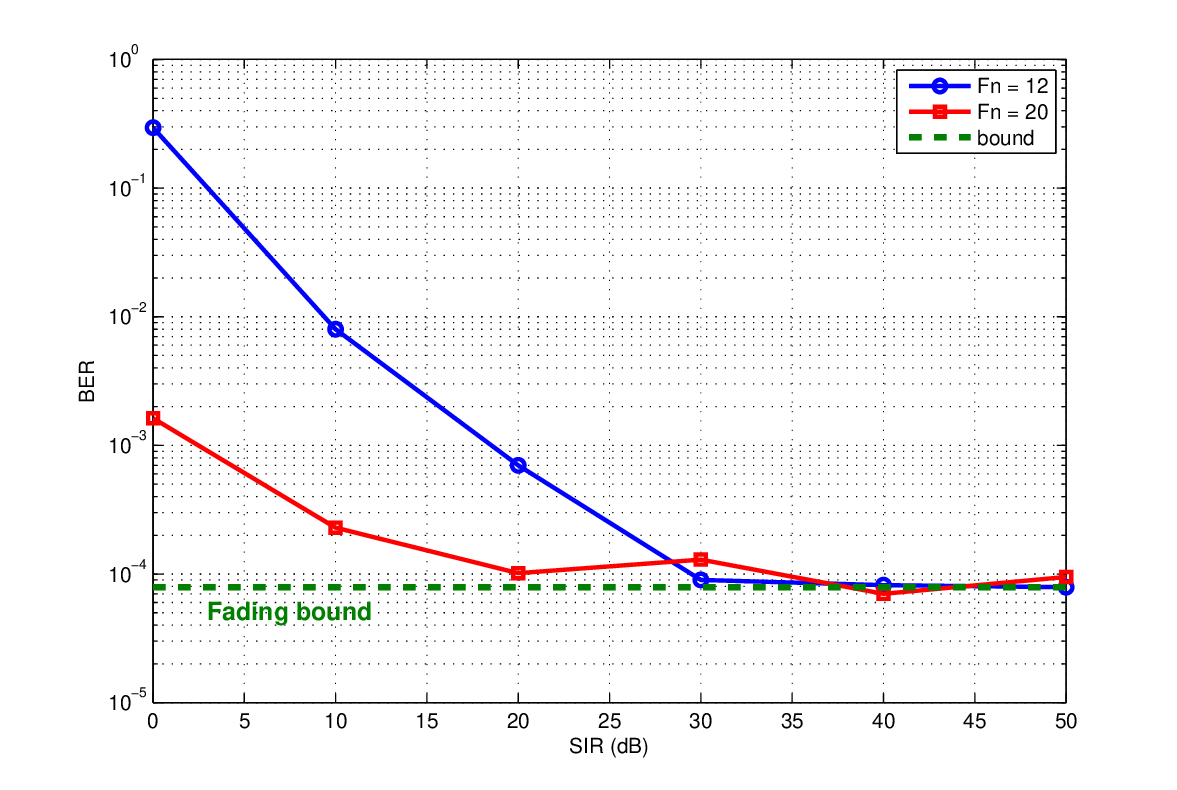}
	\caption{NB PU BER as a function of SIR in fading channel at $\frac{E_b}{N_o} = 35 \: dB$.}
	\label{fig:fading}
\end{figure}

Fig. \ref{fig:fading} shows the effect of the OFDM CU location and SIR on the NB PU BER in fading channels. One can see that the NB BER decreases as $F_n$ increases at low SIR, while as the SIR increases, this approaches an asymptotic value given by $\frac{1}{2}\left ( 1-\sqrt{\frac{E_b/N_o}{E_b/N_o +1}} \right )$ \cite{proakis2008digital}. To meet the target BER, a separation of $12\Delta F$ can be achieved for SIR = 30 dB, while this separation increases to $20\Delta F$ for SIR = 20 dB. Clearly, higher SIR and frequency separation is needed to achieve the target BER when compared with AWGN conditions.

\subsection{Coexistence between two OFDM systems}
\begin{figure}[!t]
	\centering
		\includegraphics[width=0.50\textwidth]{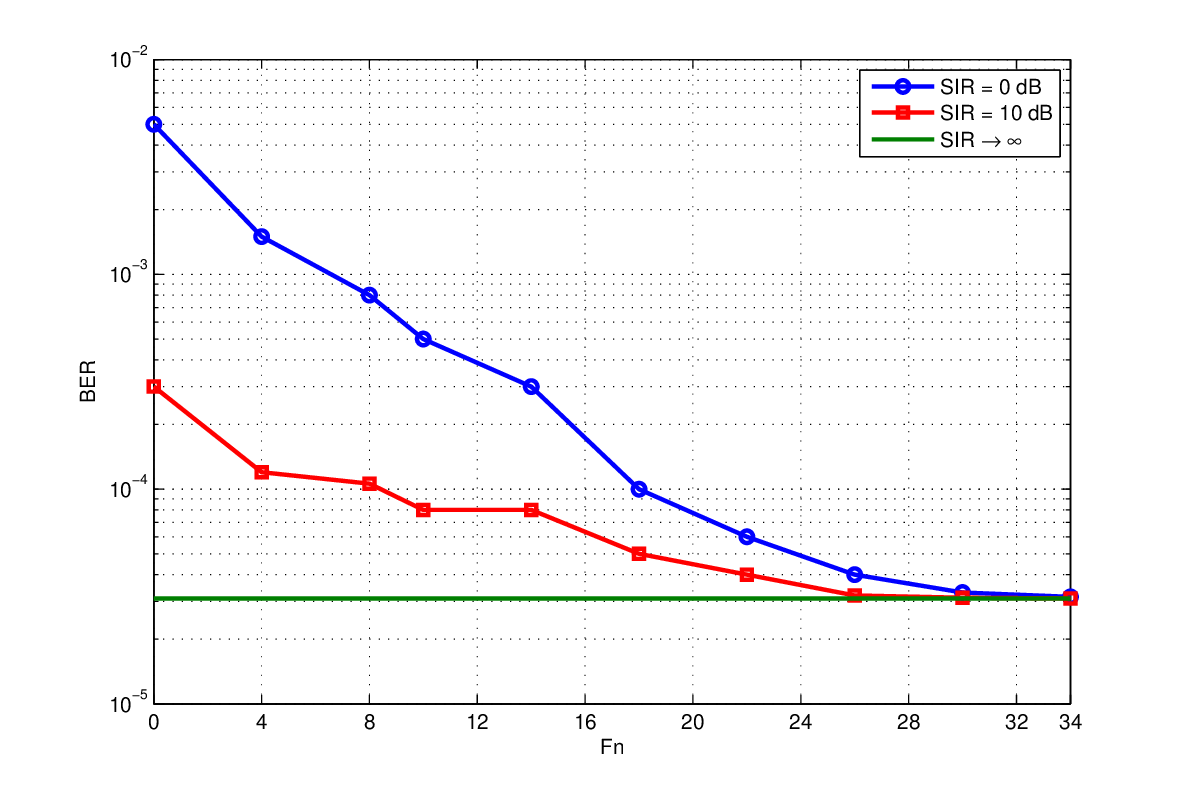}
	\caption{OFDM PU BER as a function of $F_n$ in AWGN channel at $\frac{E_b}{N_o} = 10 \: dB$.}
	\label{fig:Results_OFDM_OFDM_vs_F}
\end{figure}

For the coexistence between two OFDM systems, Fig. \ref{fig:Results_OFDM_OFDM_vs_F} shows the effect of the OFDM CU location and SIR on the OFDM PU BER in AWGN channel. The previous discussion applies and we can easily determine the minimum frequency separation to be $18\Delta F$ for SIR = 0 dB and $8\Delta F$ for SIR = 10 dB.

\begin{figure}[!t]
	\centering
		\includegraphics[width=0.50\textwidth]{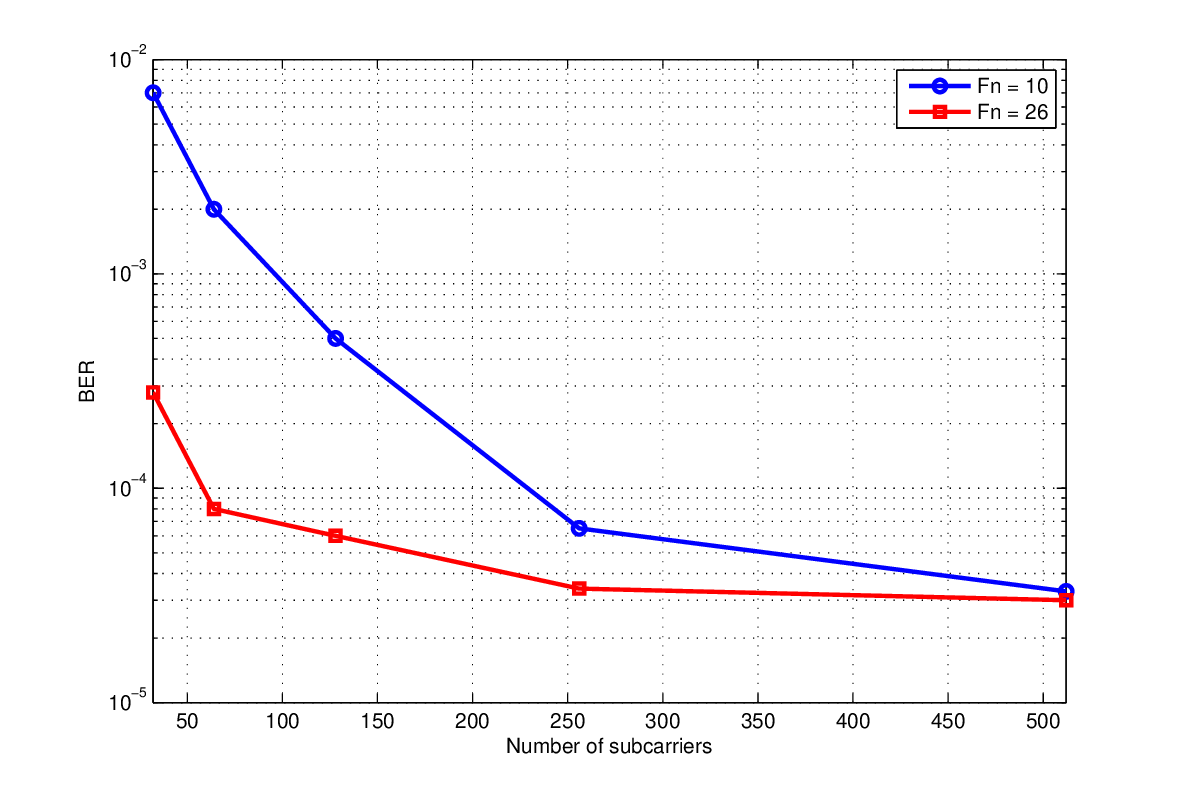}
	\caption{OFDM PU BER as a function of OFDM CU number of subcarriers in AWGN channel at SIR = 0 dB and $\frac{E_b}{N_o} = 10 \: dB$.}
	\label{fig:Results_OFDM_OFDM_vs_N}
\end{figure}

The effect of the OFDM CU number of subcarriers on the OFDM PU BER is depicted in Fig. \ref{fig:Results_OFDM_OFDM_vs_N} for different $F_n$ values. As one can notice, the OFDM PU BER decreases with increasing the CU number of subcarriers, which leads to a reduction in the minimum frequency separation. For example, the target BER is attained with 64 subcarriers at $F_n$ = 26, whereas $F_n$ = 10 is sufficient if 256 subcarriers are used.

\subsection{Effect of mitigation techniques}
\begin{figure}[!t]
	\centering
		\includegraphics[width=0.50\textwidth]{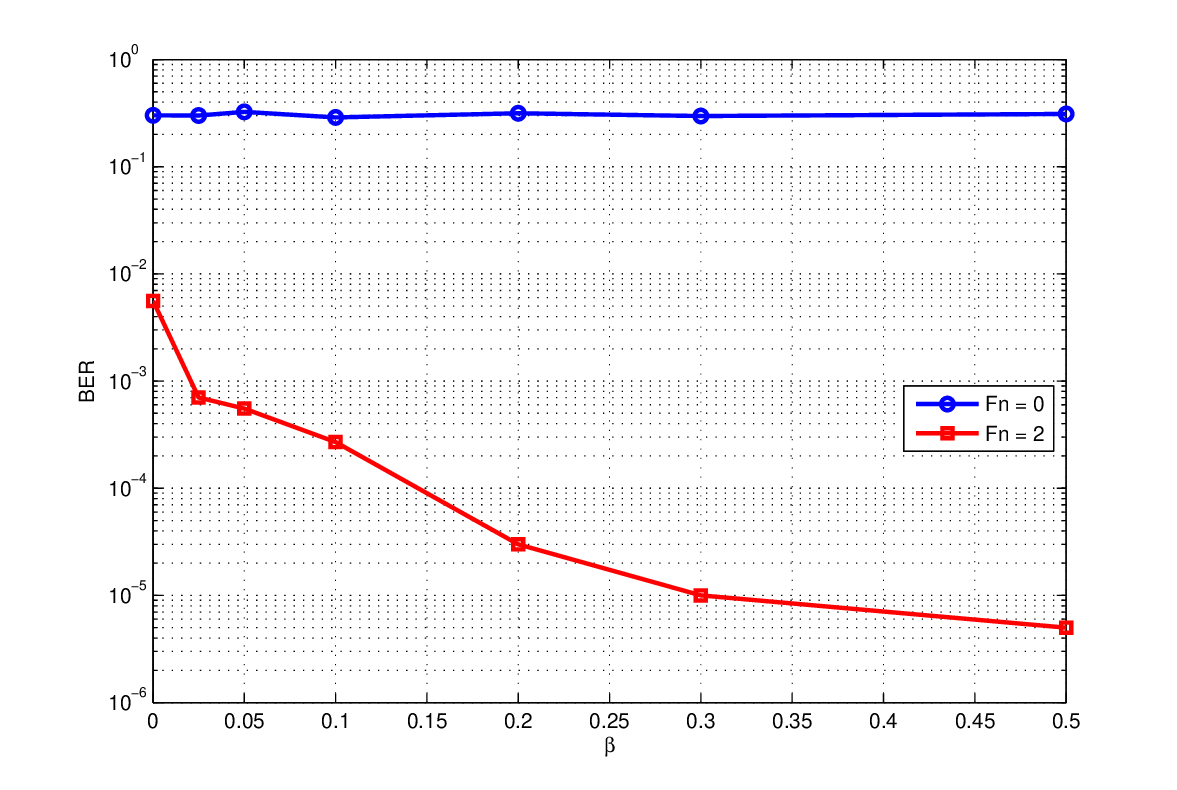}
	\caption{NB PU BER as a function of the  raised cosine window roll-off factor $\beta$ in AWGN channel at SIR = 0 dB and $\frac{E_b}{N_o} = 10 \: dB$.}
	\label{fig:Results_F_10_SIR_0}
\end{figure}

The effect of windowing on the NB PU system is investigated in Fig.  \ref{fig:Results_F_10_SIR_0} as a function of the raised cosine window roll-off factor $\beta$ for different values of $F_n$, $\frac{E_b}{N_o}$ = 10 dB, and SIR = 0 dB.  At $F_n = 0$, increasing $\beta$ has no effect on the NB PU BER, as the reduction of the OFDM side lobes occurs outside the NB PU bandwidth. At $F_n$ = 2, increasing the value of $\beta$ improves the NB PU BER, as the reduction in the OFDM side lobes occurs within the NB PU bandwidth. Hence, windowing can allow less minimum separation at the expense of increasing the OFDM CU symbol duration. For example, it was show earlier that the minimum separation for the coexistence scenario between OFDM and NB is $10\Delta F$ at SIR = 0 dB to reach the NB target BER. However, with the help of windowing, we can meet the target BER at a distance of $2\Delta F$ with $\beta$ = 0.15.

\begin{figure}[!t]
	\centering
		\includegraphics[width=0.50\textwidth]{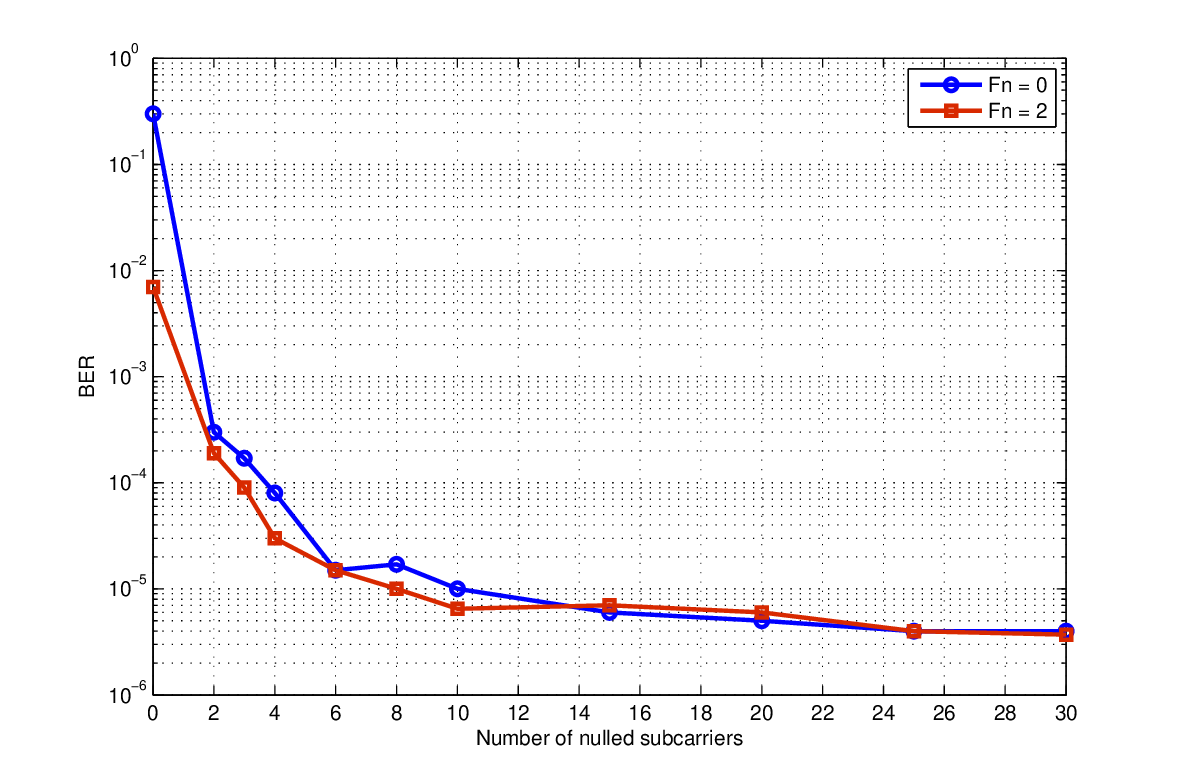}
	\caption{NB PU BER as a function of the OFDM CU number of nulled subcarriers in AWGN channel at SIR = 0 dB and $\frac{E_b}{N_o} = 10 \: dB$.}
	\label{fig:Results_Final_NB_OFDM}
\end{figure}

The effect of nulling OFDM CU subcarriers located near to the NB PU is introduced in Fig. \ref{fig:Results_Final_NB_OFDM} for different $F_n$, at SIR = 0 dB and $\frac{E_b}{N_o}$ = 10 dB. It is clear that, at $F_n$ = 0, the NB PU BER performance greatly improves when nulling a certain number of subcarriers (four subcarriers need to be nulled to reach the target BER). For $F_n$ = 2, three subcarriers should be nulled to achieve the same performance. 

\begin{figure}[!t]
	\centering
		\includegraphics[width=0.50\textwidth]{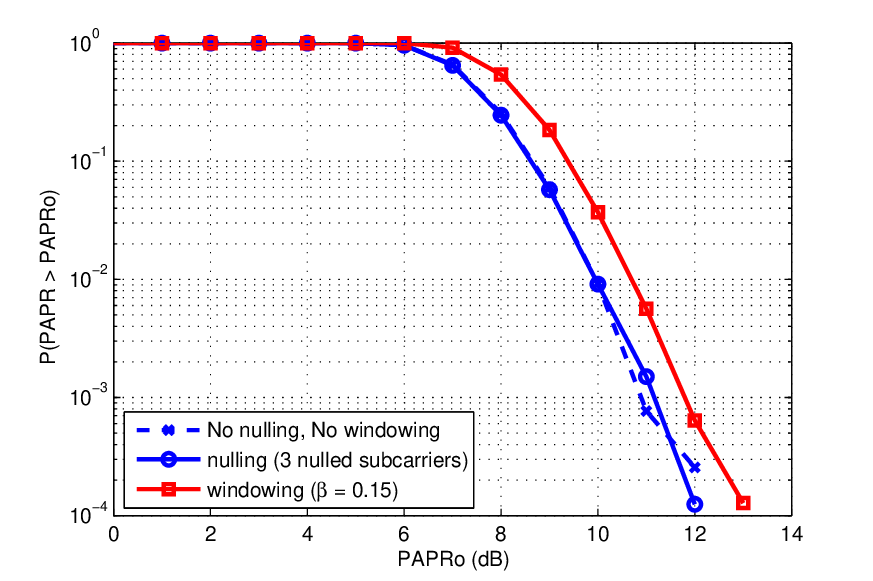}
	\caption{OFDM CU PAPR for $\beta$ = 0.15 and 3 nulled subcarriers.}
	\label{fig:PAPR}
\end{figure}

To meet the target BER in case of coexistence scenario between NB PU and OFDM CU, $\zeta$ equal to 2.7826 (bit/sec)/Hz for windowing with $\beta$ = 0.15, and 3.125 (bit/sec)/Hz in case of nulling three subcarriers (assuming QPSK modulation for both cases, i.e. $m$ = 2). This shows an advantage of using nulling over windowing in terms of spectral efficiency for the chosen parameters. In Fig. \ref{fig:PAPR}, the effect of windowing and nulling on the CU PAPR is depicted. It is clear that nulling three subcarriers yields lower PAPR when compared with windowing ($\beta$ = 0.15). More results on PAPR show that nulling more subcarriers reduces the PAPR (approaching the single subcarrier case), while increasing the value of $\beta$ increases the PAPR (the raised cosine window has a maximum of 1, so, it has no effect on $max[s(t) s^*(t)]$, while it yields a reduction in $\mathbb{E}[s(t) s^*(t)]$, see (\ref{eq:papr}) and (\ref{eq:hh})). These results are omitted due to space limitation.

\section{Conclusion} \label{sec:conc}

The coexistence between OFDM CU, and NB PU and OFDM PU systems is considered in this paper. The minimum frequency separation to meet a certain target BER in each scenario is determined, and it is found to be a function of SIR and channel conditions. The frequency separation can be improved by using either windowing or nulling subcarriers; however, this reduces spectral efficiency. In future work, we will study how to balance such trade-offs, to optimize the spectral efficiency for the coexistence scenarios.

\section*{Acknowledgment}

This work has been supported in part by the Communications Research Centre, Canada.



%




\end{document}